\documentclass[3p,times,procedia]{elsarticle}

\usepackage{ecrc}
\usepackage{caption}

\volume{00}

\firstpage{1}

\journalname{Procedia Computer Science}

\runauth{}

\jid{procs}

\jnltitlelogo{Procedia Computer Science}

\CopyrightLine{2011}{Published by Elsevier Ltd.}

\usepackage{amssymb}

\usepackage[figuresright]{rotating}
\usepackage{amssymb}
\usepackage{graphicx}
\usepackage{float}
\usepackage{amsmath}
\usepackage{subfigure}
\usepackage{rotating}
\usepackage{multirow}
\usepackage{array}
\usepackage{algorithm}
\usepackage{algorithmic}
\usepackage{url}
\urldef{\mailsa}\path|{},|

\begin{document}

\begin{frontmatter}

\author{N. Javaid\corref{cor3}\fnref{label3}}
\ead{nadeemjavaid@yahoo.com, nadeemjavaid@comsats.edu.pk}
\ead[url]{http://www.njavaid.com}

\dochead{2013 International Workshop on Communications and Sensor Networks (ComSense-2013)}

\title{AID: An Energy Efficient Decoding Scheme for\\ LDPC Codes in Wireless Body Area Sensor Networks}

\author{O. Rehman$^{1}$, N. Alrajeh$^{2}$, Z. A. Khan$^{3}$, B. Manzoor$^{1}$, S. Ahmed$^{1, 4}$}

\address{$^{1}$COMSATS Institute of Information Technology, Islamabad, Pakistan. \\
        $^{2}$B.M.T., C.A.M.S., King Saud University, Riyadh, Saudi Arabia.\\
        $^{3}$Faculty of Engineering, Dalhousie University, Halifax, Canada.\\
        $^{4}$Abasyn University Peshawar, Pakistan.\\
}

\begin{abstract}
One of the major challenges in Wireless Body Area Networks (WBANs) is to prolong the lifetime of network. Traditional research work focuses on minimizing transmit power; however, in the case of short range communication the consumption power in decoding is significantly larger than transmit power. This paper investigates the minimization of total power consumption by reducing the decoding power consumption. For achieving a desired Bit Error Rate (BER), we introduce some fundamental results on the basis of iterative message-passing algorithms for Low Density Parity Check Code (LDPC). To reduce energy dissipation in decoder, LDPC based coded communications between sensors are considered. Moreover, we evaluate the performance of LDPC at different code rates and introduce Adaptive Iterative Decoding (AID) by exploiting threshold on the number of iterations for a certain BER $(10^{-4})$. In iterative LDPC decoding, the total energy consumption of network is reduced by $20-25 \%$.
\end{abstract}


\end{frontmatter}

\section{Background and Motivation}
\label{sec:format}
Recent research in WSNs mostly deals with the aim to maximize energy efficiency (~\cite{1}, ~\cite{2}). In this regard, some authors worked at routing layer (~\cite{3}, ~\cite{4}, ~\cite{5}) while others (~\cite{6}, ~\cite{7}, ~\cite{8}) explored MAC layer. However, our directions are more focused on different decoding schemes to achieve maximum energy efficiency. In ~\cite{9}, P. Grover \textit{et al.} investigate the perspective of interference on decoding power. They suggest that in short range communication the transmit power is smaller than decoding power and uncoded transmission requires more transmit power than coded transmission. Andrea \textit{et al.} designed LDPC decoder architecture for WSN. Different LDPC codes are considered to analyze the energy saving w.r.t un-coded communication, depending on distance, BER and Environment ~\cite{10}.
S.L.Howard \textit{et al.} calculate critical distance $d_c$ at which the decoder's energy consumption per bit is equal to transmit energy per bit. In comparison to an un-coded system, authors provide results for $d_c$ in different environments over a wide frequency range ~\cite{11}. Marcelo \textit{et al.} investigate the tradeoff between transmission and processing energy consumption in sensor nodes by employing convolutional codes. For each sensor node, authors select appropriate complexity for ECC to prolong network lifetime~\cite{12}. Z. Hajjarian \textit{et al.} define the relationship between the number of quantization bits and decoder's energy consumption using LDPC in WSN. Decoder's complexity is reduced by replacement of functional blocks with look up tables. They suggest that LDPC codes are more energy efficient than conventional and block codes. Using iterative decoding, the network lifetime is increased up to four times with regular LDPC codes ~\cite{13}. In ~\cite{14}, authors propose a packet error reduction technique for reliable communication in WBAN. Firstly, they calculate Received Signal Strength (RSS) around the human body. Secondly, based on these calculations of RSS, they calculate Packet Error Rate (PER). LT codes are considered for WBAN to reduce the PER. In ~\cite{15}, authors use rateless codes to provide an adaptive duty cycling for power management. Analytical results show that with same structure used for WSNs, upto $80$ percent of energy is saved as compared to IEEE 802.15.4 physical layer standard. Z. H. Cai \textit{et al.} propose an efficient early stopping method to reduce number of iterations for LDPC decoders. This method is very efficient at low SNR~\cite{16}.

WBAN consists of small sensors with limited battery power. In this paper, our objective is to increase the network lifetime or to minimize the energy consumption of network. Reliability is another primary requirement for communication. The level of reliable communication depends on applications and user specific constraints. Error correction schemes provide reliable communication between transmitter and receiver by reducing $P_b$. In practical situations, such as WBAN there are two ways to reduce $P_b$.\\
1. Increasing the transmit power improves the SNR:\\
For Additive White Gaussian Noise (AWGN) channel, $p(b)$ for BPSK modulation is expressed as:
\begin{equation}
p_b=Q(\sqrt{\frac{2E_b}{N_0}})
\end{equation}
where $Q$ is a scaled form of the complementary Gaussian error function.\\
2. Use complex decoder by increasing the decoding power:\\
In the case of WBAN an increase in transmit power is not sufficient because increase in transmit power damages the human tissues. Another problem is the existence of small distance between receiver and transmitter. The transmit power increases the Signal to Interference and Noise Ratio (SINR). In this case how we can improve SINR? This is done by increasing the distance between transmitters. However, in WBAN it is not possible because of limited area. In any communication system a tradeoff between transmit power and decoding power exists. WBAN requires low transmitting power so, there is need to implement energy efficient decoder to reduce the decoding power consumption to enhance the network lifetime. We use LDPC, because its performance for low BER communication is same as complex decoders such as viterbi decoding algorithms.

\vspace{-0.5cm}
\section{LDPC Codes}
\vspace{-0.3cm}
\label{sec:pagestyle}

Like all linear block codes, LDPC codes can be represented in two different possibilities. They  can be described via graphical representation and matrices. LDPC codes are represented as as Tanner Graph that contain two set of nodes: Check Nodes (CNs) and Variable Nodes (VNs). VNs are associated with N bits of codeword and CNs corresponds to M parity-check constraints. Edges in Tanner graph correspond to $1's$ in H and exchange of information along these edges as shown in Fig.2. Belief Propagation (BP) algorithm is commonly used to decode LDPC codes.



  \begin{figure}[!ht]
    \begin{minipage}{0.49\linewidth}
\includegraphics[height=9cm,width=7cm ,angle=0]{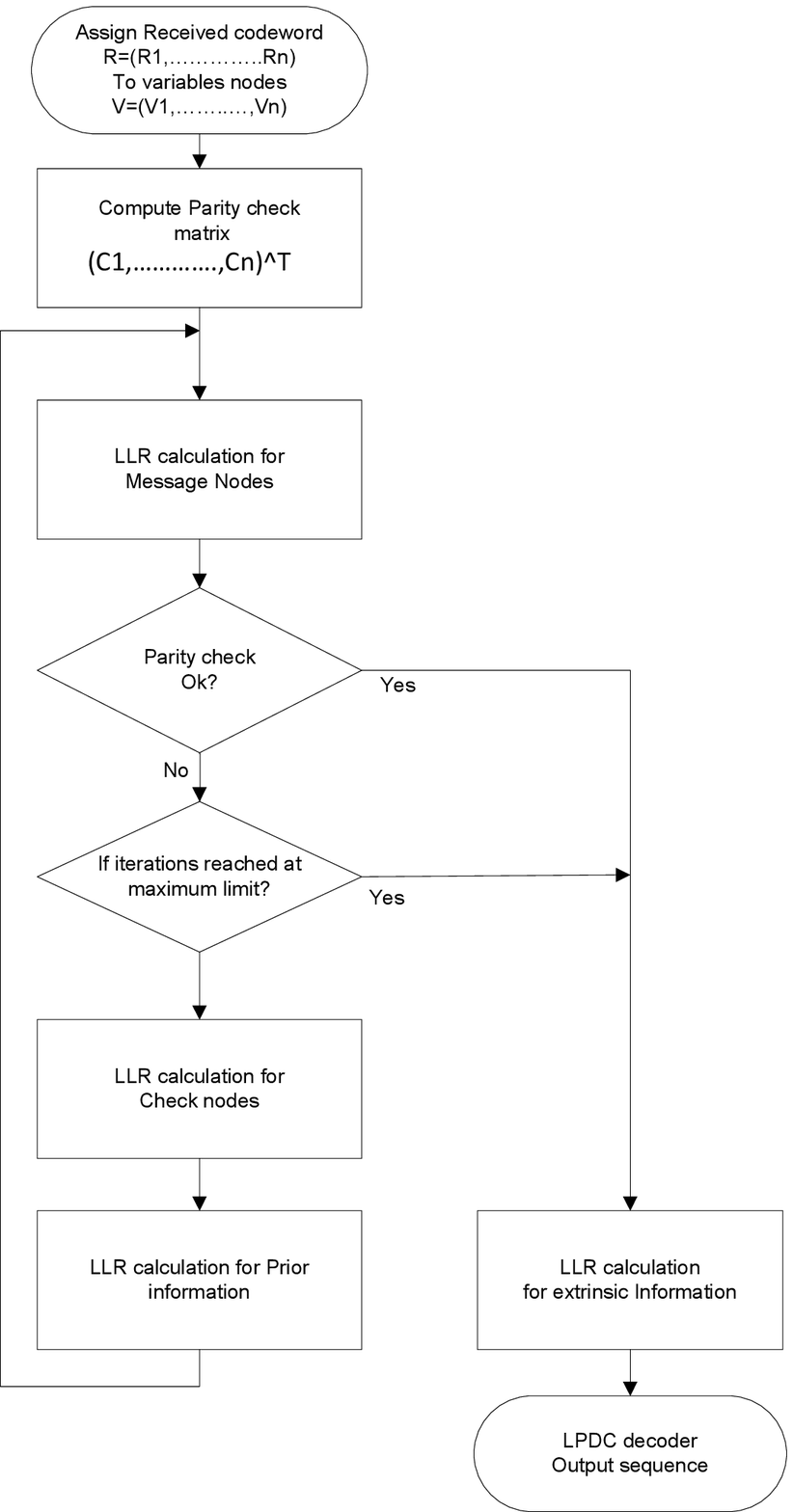}
\vspace{-0.4cm}
\caption{Flow chart of LDPC decoder}
    \end{minipage}
    \hspace{1.5cm}
    \begin{minipage}{0.49\linewidth}
\includegraphics[height=4cm,width=5cm]{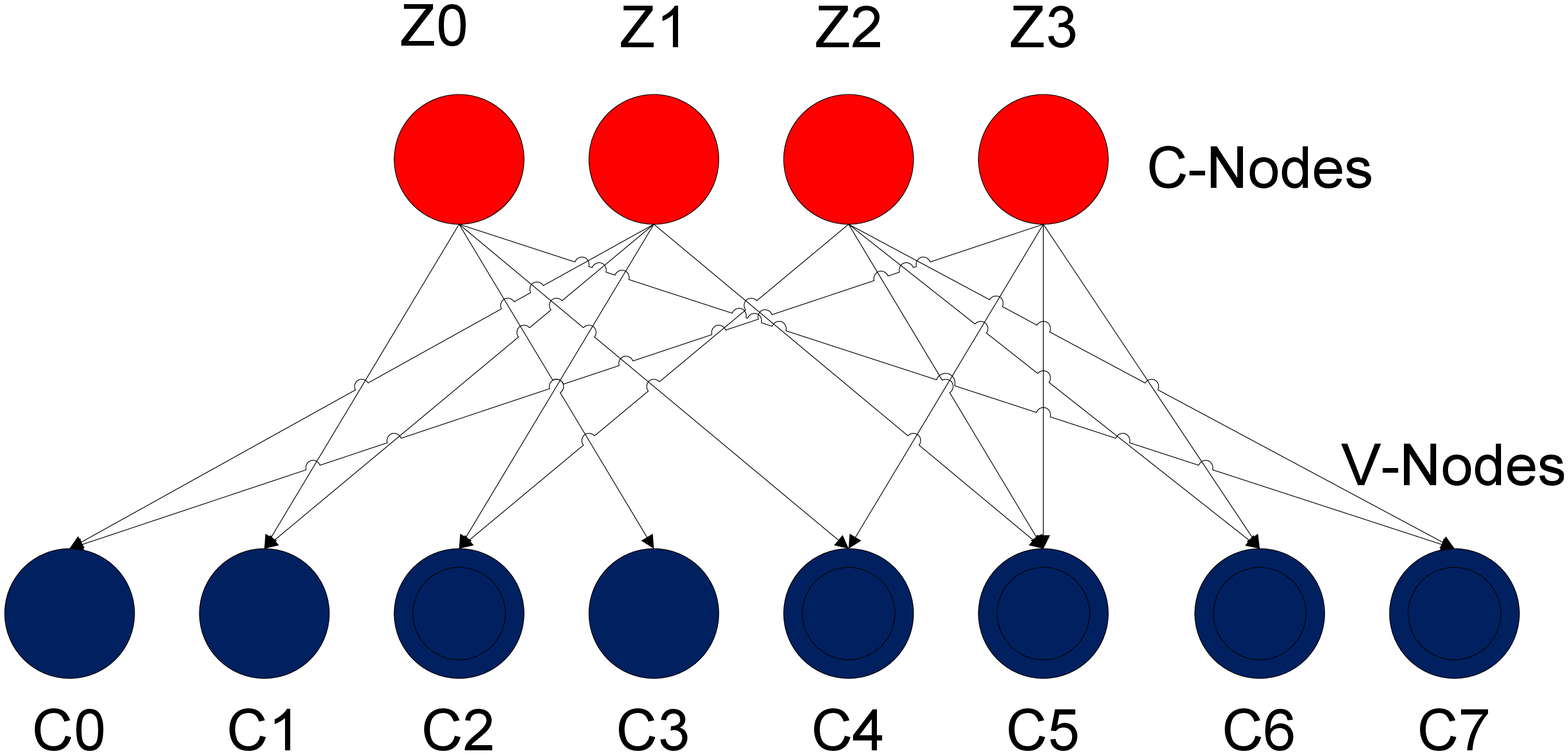}
\caption{Tanner graph for LDPC code}
    \end{minipage}
  \end{figure}


Let $S_j$ represents the LLR of the bit in column $j$ of $H$. LLR is initialized to the corresponding received soft value.
For each parity constraint $m$ in a given layer, the following operations are executed [5]:
\[ H = \left| \begin{array}{cccccccc}
0  & 1 & 0 & 1 & 1 & 0 & 0 & 1 \\
1  & 1 & 1 & 0 & 0 & 1 & 0 & 0 \\
0  & 0 & 1 & 0 & 0 & 1 & 1 & 1 \\
1  & 0 & 0 & 1 & 1 & 0 & 1 & 0 \\
\end{array} \right|.\]
\begin{equation}
Q_{mj}=S_j^{(old)}-R_{mj}^{(old)}
\end{equation}
\begin{equation}
A_{mj}=\sum_{\eta\epsilon N_m,n\neq j}\psi (Q_{mn})
\end{equation}
\begin{equation}
s_{mj}=\sum_{\eta\epsilon N_m,n\neq j}sgn (Q_{mn})
\end{equation}
\begin{equation}
R_{mj}^{(new)}=-s_{mj}.\psi (A_{mj})
\end{equation}
\begin{equation}
S_j^{(new)}=Q_{mj}+R_{mj}^{(new)}
\end{equation}


$S_j^{(old)}$ is the extrinsic information received from previous layer and updated in equation (5). This information is propagated to succeeding layer. $R_{mj}^{(old)}$ is used to compute equation (1) then updated in equation (4). Term, $R_{mj}^{(new)}$ is used again in following iteration. $N_m$ in equations  (2) and (3) represents set of all bits connected to parity constraint m. Whole operation of LDPC decoder is given in Fig.1.

\vspace{-0.5cm}
\section{Sensor Node Energy Model in WSN}
\label{sec:pagestyle}
\vspace{-0.3cm}

For encoded data transmission, the energy consumption per bit of a node is the sum of radio energy per bit $E_r$ and computation energy per bit $E_{comp}$.
Radio energy is sum of SLEEP, TRANSIENT, and ON mode energies. The transceiver is off all the times and keeping it ON only when data is
transmitting or receiving. The radio energy per bit for transmitting $N$ bits is calculated  as in [12]:
\vspace{-0.3cm}
\begin{equation}
E_r=\frac{p_{sleep}T_{on}+p_{tran}T_{tr}+p_{on}T_{on}}{N}
\end{equation}

Power consumed in radio during ON-mode is the sum of transmitting power $p_t$ and power consumption in circuitry $p_{ckt}$ .
The radio energy consumption is:
\vspace{-0.5cm}
\begin{equation}
E_r=\frac{(p_t+p_{ckt})T_{on}}{N}
\end{equation}
Transmission power required in free space is expressed by using Friis equation:
\vspace{-0.5cm}
\begin{equation}
p_t=\frac{p_r}{G_rG_t}\times(\frac{4\pi}{\lambda})^2d^n
\end{equation}

The received power is calculated as:
\vspace{-0.5cm}
\begin{equation}
p_r=SNR_{uncod}bB\frac{N_0}{2}NF
\end{equation}
where, $SNR_{uncod}$ is SNR for transmitting uncoded data, $b$ is number of bits per modulation symbol, $B$ is bandwidth ,$\frac{N_0}{2}$
is noise spectral density and $NF$ is Noise Figure.

For coded data received power is:
\vspace{-0.5cm}
\begin{equation}
p_r=SNR_{coded}bB\frac{N_0}{2}NF
\end{equation}
Transmission energy for coded data frame is expressed as:
\vspace{-0.4cm}
\begin{equation}
E_{trans}^{frame}=E_{dec}^{bit}.f_{size}^C
\end{equation}
The required processing energy to decode a bit is $E_{dec}^{bit}$. Therefore, the energy required to decode a frame $ E_{dec}^{frame}$ is computed as:
\vspace{-0.5cm}
\begin{equation}
E_{dec}^{frame}=E_{dec}^{bit}.f_{size}^C.r (J/F)
\end{equation}
where $r$ is the coding rate.

In [9] the shannon capacity limit for both coded and uncoded systems is shown as in Fig. 3.

  \begin{figure}[!ht]
    \begin{minipage}{0.49\linewidth}
\includegraphics[height=4cm,width=5cm,angle=0]{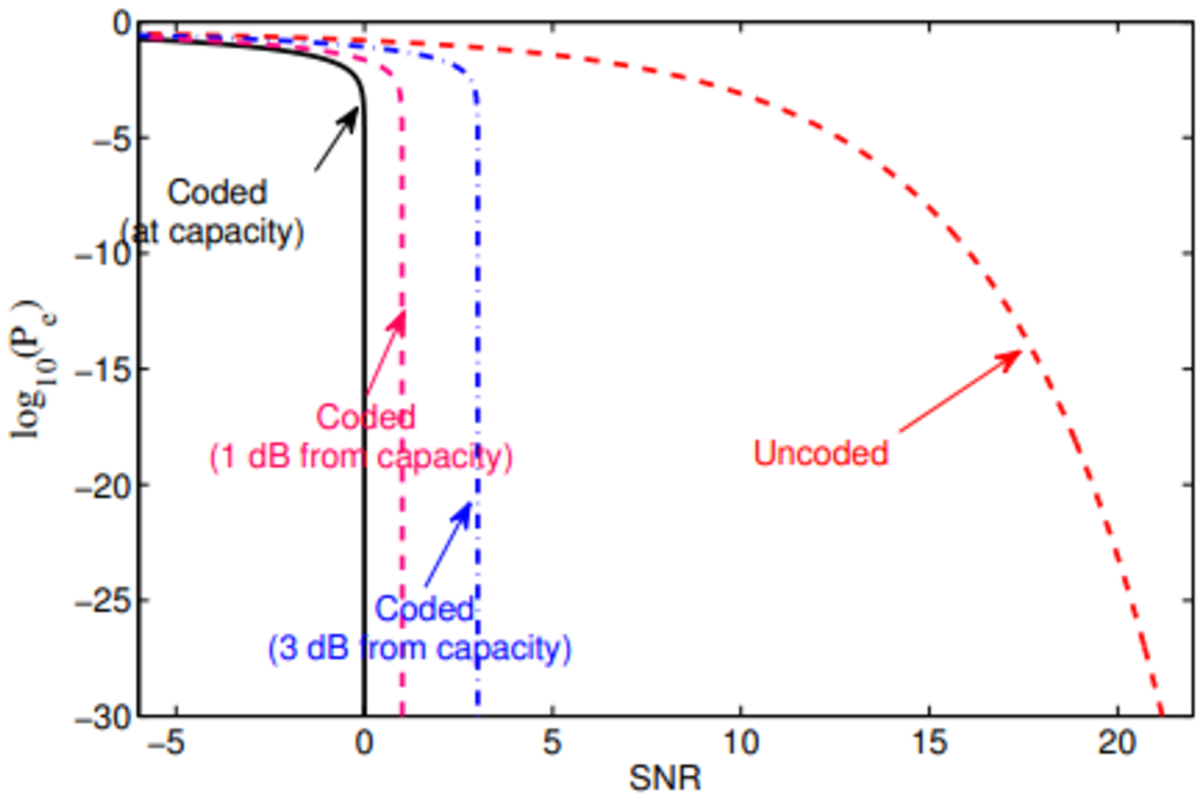}
\caption{Shannon Capacity Estimation}
    \end{minipage}
    \hspace{0.1cm}
    \begin{minipage}{0.49\linewidth}
\includegraphics[height=10cm,width=7cm ,angle=0]{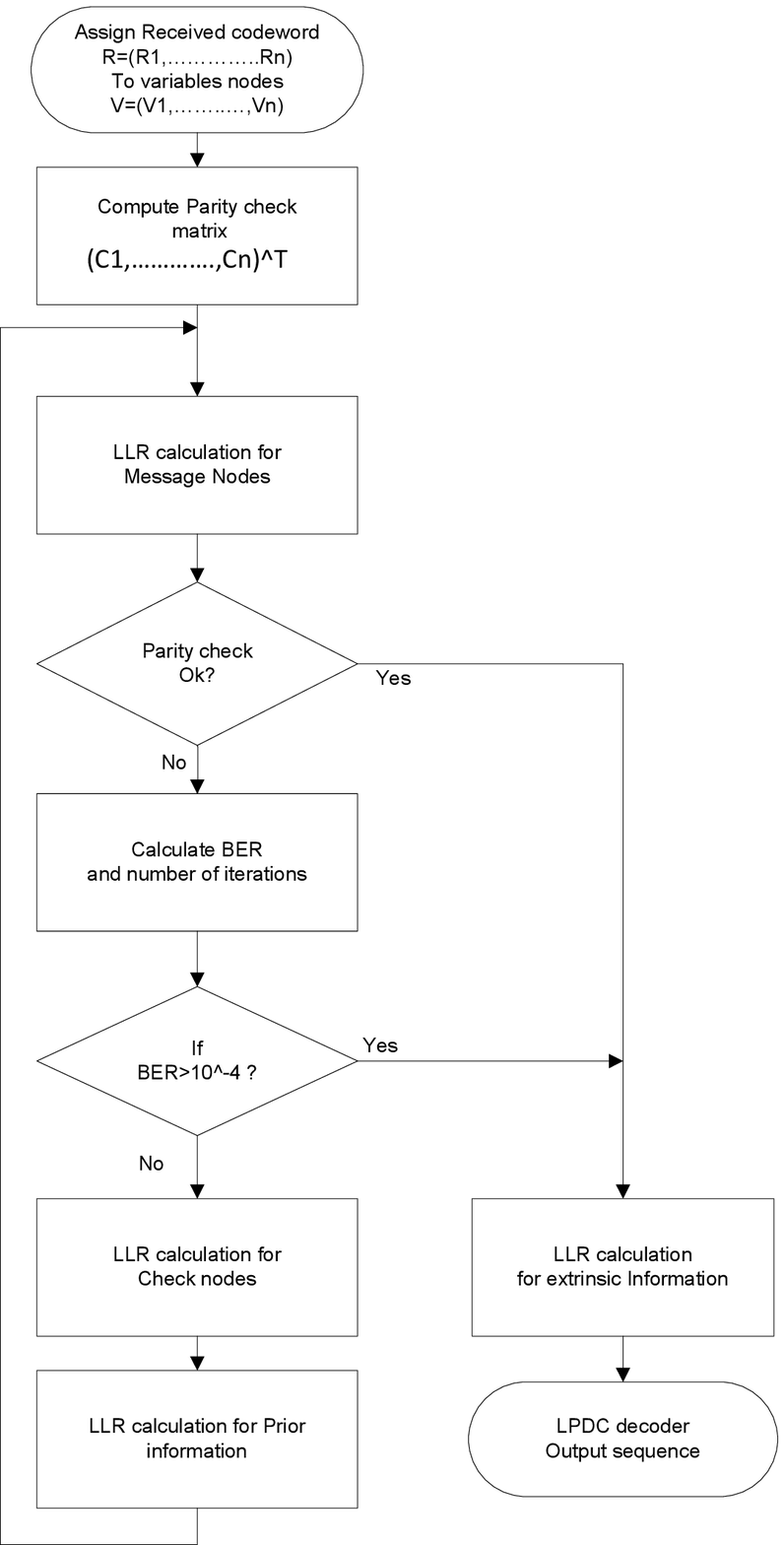}
\vspace{-0.5cm}
\caption{Flow chart of AID for LDPC}
    \end{minipage}
  \end{figure}

\vspace{-1.3cm}
\section{AID}
\vspace{-0.3cm}

 AID predicts the threshold at number of iterations for undecodable blocks. Another advantage is reduction in processing delay. LDPC codes yield excellent performance when they are decoded with iterative message passing algorithms. The performance of an iteratively decoded LDPC code is typically reported in terms of $p_e$ for a specific SNR value. For high SNR values, the value $p_e$ is small that requires a large number of iterations in order to obtain reliable estimate. In AID, we introduce a threshold on the number of iterations after achieving a certain BER $(10^{-4})$. The whole process of AID is shown in Fig.4.

We assume that each PE consumes a fixed $E_{node}$ joules of energy per iteration. We assume that the $ R_{dec}$ is equal to data rate $R_{data}$  measured in information bits per second.
The power received at distance $x$ meter is given in Eq [9].
In Eq [13] the $E_{dec}$ is the energy consumed in decoder operation.
For any rate $R_{data}$ the average probability of error $(p_e)\longrightarrow 0$.
Let the number of decoding iterations be denoted by $l$, the number of computational nodes can be lower bounded by m, the number of received channel outputs.
Since each node consumes $E_{node}$ joules of energy in each iteration , the decoding energy $E_{dec}$ is lower bounded in [2] as:
$
E_{dec}\geq E_{node}\times m \times l
$. We assume channel encoding is free. This result is the following lower bounds on the weighted total power is $E_{tot}\geq p_T+\frac{E_{node}\times m \times l}{T_{dec}}$, and $E_{tot}=P_LP_R+\frac{E_{node}\times m \times l}{T_{dec}}$. Where $T_{dec}=\frac{k}{R_{dec}}$ is the time consumed in decoding. Thus,
$p_{total}\geq P_LP_R+\frac{E_{node}\times m \times l \times R_{dec}}{k}$, and $p_{total}=P_LP_R+\frac{E_{node} \times l \times R_{dec}}{R_{c}}$, where, $P_L$ is power loss and $P_R$ is received power.

  \begin{figure}[!ht]
    \begin{minipage}{0.49\linewidth}
    \includegraphics[height=5cm,width=7cm]{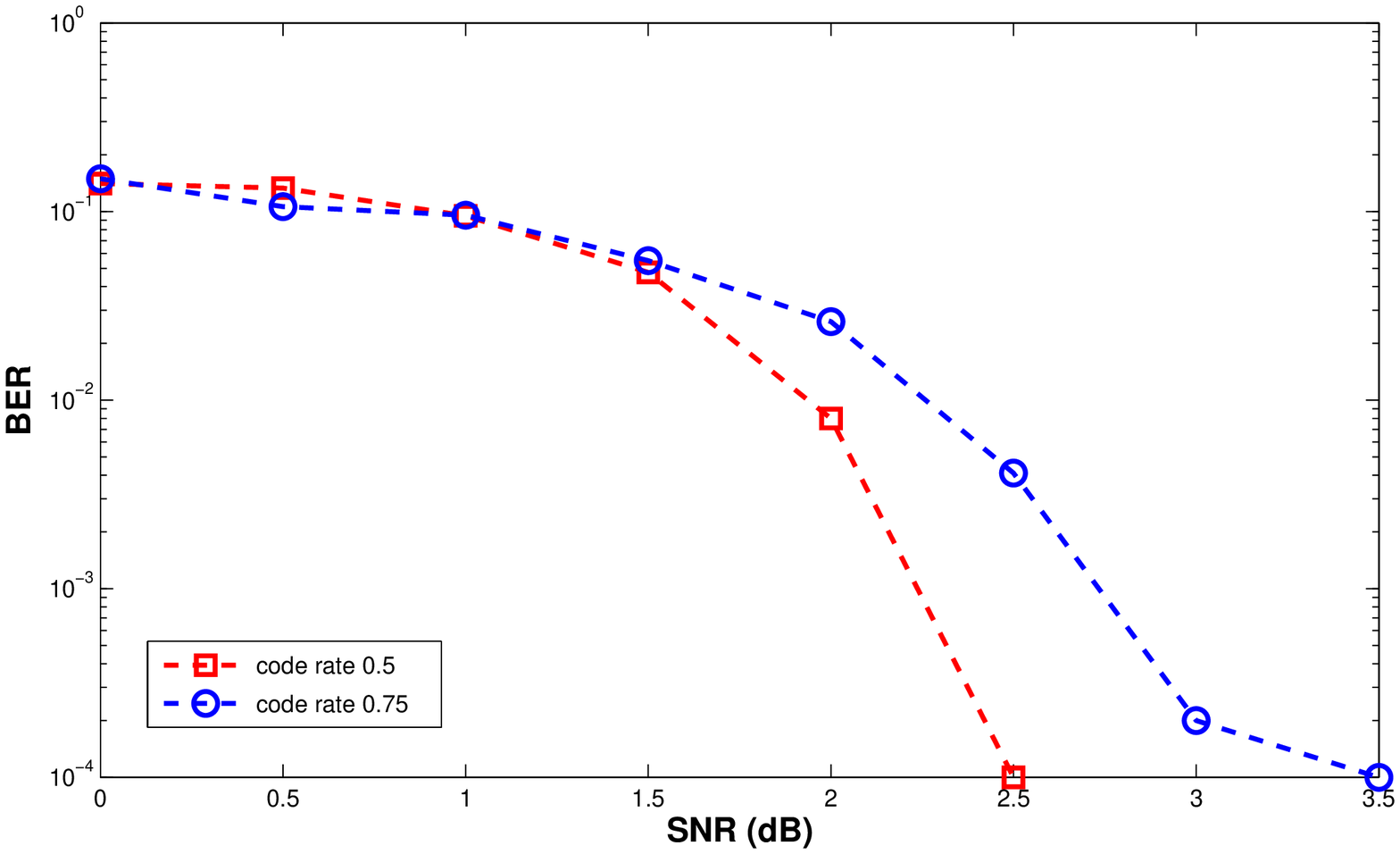}
    \vspace{-0.5cm}
    \caption{Relationship between SNR and BER using LDPC}
    \end{minipage}
    \begin{minipage}{0.49\linewidth}
    \includegraphics[height=5cm,width=7cm]{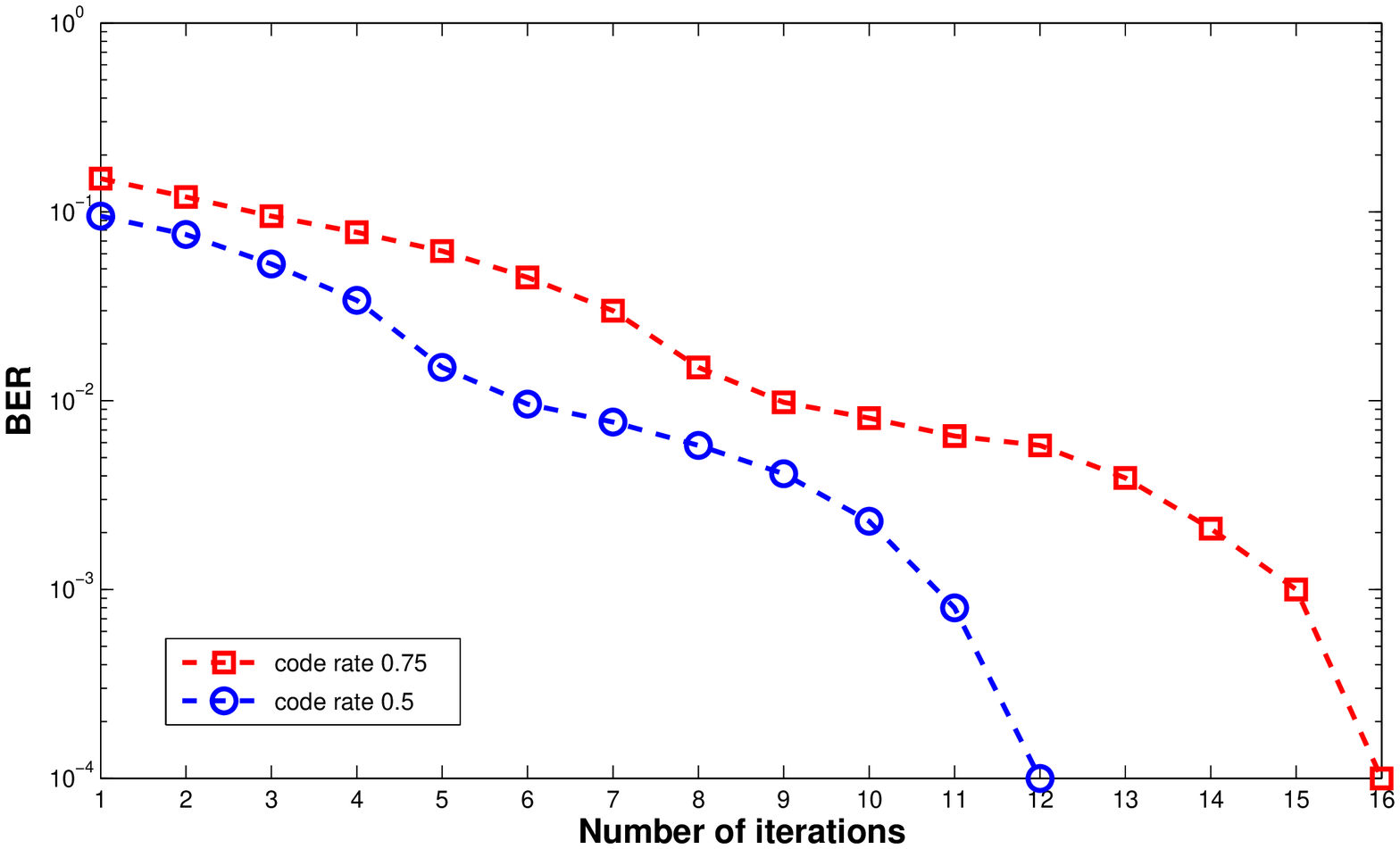}
    \vspace{-0.5cm}
    \caption{Iterations in AID for desired BER( $10^{-4}$ )}
    \end{minipage}
     \begin{minipage}{0.49\linewidth}
    \includegraphics[height=5cm,width=7cm]{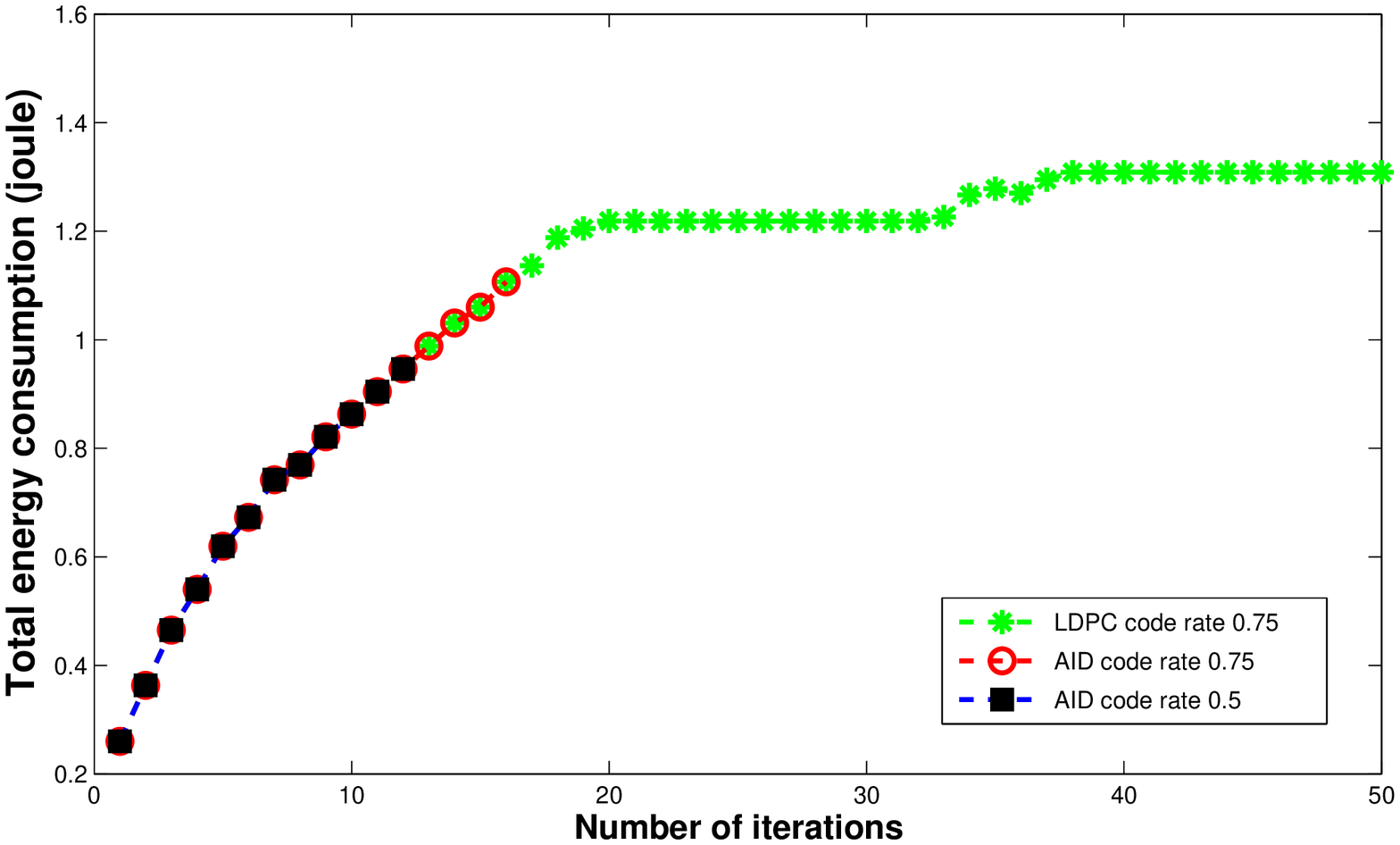}
    \vspace{-0.5cm}
    \caption{Energy Consumption in AID}
    \end{minipage}
    \hspace{1cm}
    \begin{minipage}{0.49\linewidth}
    \centering
    \scriptsize
    \begin{tabular}{|c|c|}
\hline
       PTr &      144mW \\
\hline

         w &          1 \\
\hline

pathloss Exonent &          1 \\
\hline

         d &             3m \\
\hline

Boltzmann Constant & $1.38065*10^-23$ \\
\hline

temperature &        300 \\
\hline

     Rdata &    1.5Gbps \\
\hline

      Rdec &    1.5Gbps \\
\hline

     Enode &     $10^-12 $\\
\hline

    degree &          2 \\
\hline

        Rc &        0.5 \\
\hline

      LDPC &        7,4 \\
\hline

 frequency &     60 GHZ \\
\hline
\end{tabular}

\setlength{\hoffset}{-2cm}
\caption{Simulation Parameters}

    \end{minipage}
  \end{figure}


%

\vspace{-1cm}
\section{Results and Discussions}
\label{sec:typestyle}
\vspace{-0.3cm}

LDPC code with Code Rate $(R_c)$ has been used in simulation over AWGN channel with an iterative probabilistic decoding algorithm.
The bandwidth is large (3GHZ) and the throughput is 1.5 Gbps. The power of a PE is 10 pico-Joules. For indoor environment, path-loss exponent are assumed to be 3 and noise figure is 3dB. The transmitter power is in few milli Watts for 3m distance between sensor and base station. If the transmit power $P_T$ is extremely close to that required for channel capacity then large number of iterations $l$ are required. Large number of iterations consume high decoding power. Encoder should have larger power as compared to its Shannon limit in order to minimize decoding power consumption. Simulations parameters are shown in Fig.8.



Fig.5, shows the relationship between the SNR and BER, as we increase the SNR, BER decreses. By imposing five iterations $(l=5)$, it is observed that targeting a specific BER of $10^{-4}$ the requried SNR is $3.5$ at $R_c$ 0.75 and SNR 2.5 at $R_c$ 0.5.

Fig.6, shows the relationship between number of iterations and BER. For a fixed SNR (1.5dB) our target is to achieve a  BER of $10^{-4}$ and calculate the number of iterations till desired BER of $10^{-4}$ is achieved. Results show that at $R_c$, 0.75 the required iterations are 16 and at $R_c$ 0.5, 12 iterations are required.

Fig.7, shows the relationship among the total number of iterations and total power consumption by using LDPC decoder and AID. For the case of LDPC code, we use 50 iterations at a code rate of 0.75. There is no threshold for achieving a certain value of $Pe$. As $Pe \rightarrow 0$ the total power consumption $p_T \rightarrow \infty$. Simulation results show that $P_T$ increases by increasing the number of iterations. For 50 iterations $P_T$ is 1.3 joule.
In AID, our requirement is to achieve a BER of $10^{-4}$, it is achieved at 16th iteration. Hence, we save the energy consumption of 35 iterations and also minimize the processing delay of decoder. This scheme is equally performed well for such applications in which we can not afford delay. Simulation results show that the total energy consumption of error correction code is exactly 1.05 joule. With the code rate of 0.5, energy consumed is 0.5 joule in total. We save 22$\%$ of total energy as compared to LDPC codes.


%
%
%
%
%
%
%
%
%
%
%
%
%
%
%
%
%
%

\vspace{-0.4cm}
\section{Conclusion}
\label{sec:majhead}
\vspace{-0.3cm}
In this paper, we proposed AID Scheme to reduce the decoding power consumption and to prolong the lifetime of WBANs. To achieve BER to $(10^{-4})$, we calculate the total number of iterations required for LDPC decoder. As $p_b$ approaches $0$, iterations approach infinity and iterations stop after achieving a desired BER. We reduce decoding energy consumption by stoping the iterations after achieving a certain BER. By using this scheme, $20-25\%$ of total energy consumption is reduced.

\vspace{-0.4cm}
\section{Acknowledgment}
\vspace{-0.3cm}
The authors extend their appreciation to the Distinguished Scientist Fellowship Program (DSFP) at King Saud University
for funding this research.

\end{document}